\documentstyle[psfig]{l-aa}
\begin{document}

\thesaurus{ 08.01.2, 08.02.1, 08.03.3, 08.12.1, 08.09.2 }

\title{Li~{\sc i} enhancement during a long-duration 
stellar flare}

\author{David Montes \inst{1,}\inst{2,}\inst{3} 
\and  Lawrence W. Ramsey \inst{2,}\inst{3} 
}

\offprints{David Montes}

\institute{
Departamento de Astrof\'{\i}sica,
Facultad de F\'{\i}sicas,
 Universidad Complutense de Madrid, E-28040 Madrid, Spain
(dmg@astrax.fis.ucm.es)
\and
The Pennsylvania State University,
Dep. of Astronomy and Astrophysics,
525 Davey Laboratory, University Park, PA 16802
\and
Guest observer at McDonald Observatory}

\date{Received ; accepted }

\maketitle

%**************************
\begin{abstract}
We report the possible detection of a Li~{\sc i} $\lambda$6708 \AA\
line enhancement during an unusual long-duration optical flare in
the recently discovered,
X-ray/EUV selected, chromospherically active binary 2RE~J0743+224.
The Li~{\sc i} equivalent width (EW) variations 
follow the temporal evolution of the flare
and large changes are observed in the intensity of the line.
The maximum Li~{\sc i} enhancement (40\% in EW) occurs just after
the maximum chromospheric emission observed in the flare.
A significant increase of the $^6$Li/$^7$Li isotopic ratio is also detected.
No significant simultaneous variations are detected in other photospheric lines.
Neither line blends nor starspots
seem to be the primary cause
of the  observed Li~{\sc i} line variation.
From all this we suggest that this Li~{\sc i} enhancement
is produced by spallation reactions during the flare.


\end{abstract}

%----------------------------------------------------------------
\section{Introduction}
%----------------------------------------------------------------

The resonance doublet of  Li~{\sc i} at $\lambda$6708 \AA\
is an important diagnostic of age in late-type stars
since it is easy destroyed by thermonuclear reactions in the 
stellar interiors.
However, Li~{\sc i} observations of several types of chromospherically 
active stars, such as pre-main sequence stars, late-type dwarfs in open 
clusters, as well as post-main sequence objects, 
show a significant range of abundances.
It is well-known that a large number of
chromospherically active binaries (CAB hereafter)
show Li~{\sc i} abundances higher than
the normal values characteristic of stars of the same mass and evolutionary
stage
(Pallavicini et al. 1992, Liu et al. 1993; 
Fern\'{a}ndez-Figueroa et al. 1993;
Randich et al. 1994; 
Barrado et al. 1997, 1998; Montes et al. 1998).
This line is very temperature sensitive because of its
low ionization potential of 5.37~eV. 
Thus the equivalent width (EW) should be enhanced in dark
spots but reduced in the bright facular regions as 
demonstrated by solar observations (Giampapa 1984).
A number of researchers 
(Fekel 1996; 
Barrado 1996; Hussain et al. 1997; 
Fern\'{a}ndez \& Miranda 1998; Neuh\"auser et al. 1998)
have investigated starspots as the possible cause
of the observed Li~{\sc i} abundances spreads.
However, Li can also be produced by low energy spallation reactions 
in stellar flares (Fowler et al. 1955; Canal 1974; 
Canal et al. 1975).
Some evidence of Li production by spallation have been found 
in the Sun (Livshits 1997; Livingston et al. 1997), 
but no detection has been reported in other stars.
Spallation has only been 
discussed as a possibility to explain the high Li abundances observed in
some stars (Pallavicini et al. 1992; Mathioudakis et al. 1995;
Favata et al. 1996).

In this letter we describe the Li~{\sc i} $\lambda$6708 \AA\
line enhancement detected during the observations 
of an unusual long-duration optical flare in 
the recently discovered,
X-ray/EUV selected CAB 2RE~J0743+224,  
and discuss the possible causes.

%\placefigure{fig:rej0743_li}
%%+++++++++++++++++++
\begin{figure}
{\psfig{figure=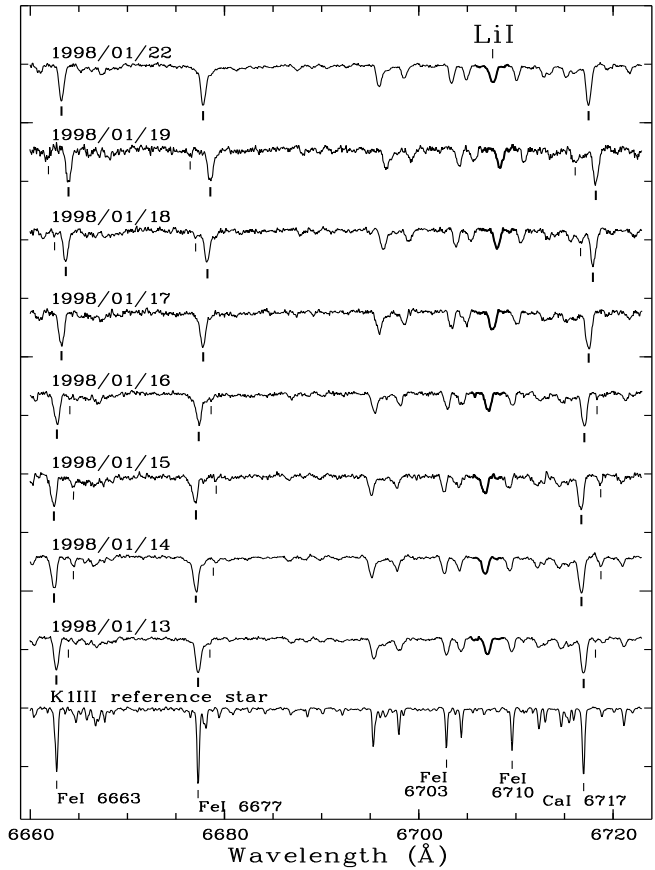,bbllx=57pt,bblly=28pt,bburx=514pt,bbury=776pt,height=11.5cm,width=8.65cm,clip=}}
\caption[ ]{Spectra of 2RE~J0743+224 in the Li~{\sc i} $\lambda$6708~\AA\
line region from 1998 January 13 to 22.
A K1~III reference star (HR 5340) is also plotted for comparison, with the
more intense photospheric lines identified.
The wavelength position of
the lines of both components are marked.
\label{fig:rej0743_li} }
\end{figure}
%%+++++++++++++++++++

%.................................................
\section{Observations and analysis}
%.................................................

High resolution (0.16~\AA) observations of this star were obtained
during a 10 night run 1998 January 12-21 using the 2.1m telescope at
McDonald Observatory and the Sandiford Cassegrain Echelle Spectrograph 
(McCarthy et al. 1993).
The spectra in the Li~{\sc i} line region are plotted in 
Fig.~\ref{fig:rej0743_li}.
These observations are analyzed in detail in a separate paper
(Montes \& Ramsey 1998b) where stellar properties, orbital parameters, and 
chromospheric behavior are discussed.
We found that this star, previously 
classified as single-lined spectroscopic binary 
(Jeffries et al. 1995),
is a double-lined spectroscopic binary (SB2)
with a K1~III primary and an orbital period of 10 days.

\begin{figure}
{\psfig{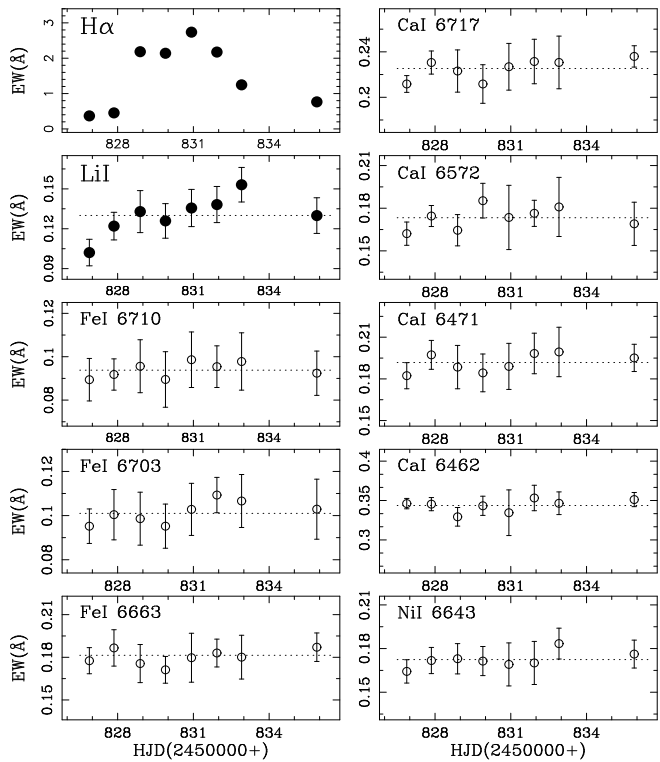}}
\caption[ ]{Measured EW for H$\alpha$, Li~{\sc i} 6708\AA\ and other
photospheric lines during the flare.
The dotted line is the mean EW for each line.
\label{fig:li_ews} }
\end{figure}

A dramatic increase in the chromospheric emissions
(H$\alpha$ and Ca~{\sc ii} IRT lines)
is detected during the observations (see the temporal evolution of 
the H$\alpha$ EW in the upper left panel of Fig.\ref{fig:li_ews}).
The increase of the emission start the 3th night (1998 January 15)
reach a maximum on the 5th night and at the end of the observations
(1998 January 22) the chromospheric lines had not yet recovered the
quiescent value. The total duration of the event was evidently larger 
than 8 days.
We interpret this behaviour  
as an unusual long-duration flare based on
a) the temporal evolution of the event,
b) the broad component observed in the H$\alpha$ line profile,
c) the detection of the He~{\sc i} D$_{3}$ line in emission and
d) a filled-in of the He~{\sc i} $\lambda$6678~\AA\ line.
A detailed description of the flare is given in Montes \& Ramsey (1998a, b).

Here we analyze the spectra in the Li~{\sc i} $\lambda$6708~\AA\ 
line region. 
The Li~{\sc i} absorption feature is clearly observed  
(Fig.~\ref{fig:rej0743_li}) 
and appears centered at the
wavelength corresponding to the primary component 
with no evidence for a contribution from the secondary.
The mean EW is 130~m\AA\ which is a little bit above the normal
value for this kind of star,  
typically $<$ 100~m\AA, (Randich et al. 1994; Barrado et al. 1997, 1998). 
In addition, 
a small absorption feature corresponding to the Li~{\sc i} $\lambda$6104~\AA\ 
is marginally detected in the spectrum taken the 4th night.
A careful analysis of the Li~{\sc i} $\lambda$6708~\AA\ line indicates that 
the line profile, EW, and intensity, I, are changing during the observations.
The measured EW and I are given in Table~\ref{tab:li} 
and the EW is plotted in Fig~\ref{fig:li_ews},
where we can see that the increase of Li~{\sc i} line 
follows the temporal evolution of the flare.
The maximum Li~{\sc i} enhancement (50\% in EW) occurs just after
the maximum chromospheric emission observed in the flare (5th night).
However, any real variations of the Li~{\sc i} EW 
(in excess of the uncertainties) 
are clearly indicated on nights 1 and 7 only.  

In order to test if these variations are real we have also measured the 
EW of other photospheric lines.
We have selected several isolated lines in the same spectral order
as the Li~{\sc i} line; Fe~{\sc i} $\lambda$6710.3~\AA\ 
and Fe~{\sc i} $\lambda$6703.6~\AA\ 
which are close and similar in strength to the Li~{\sc i} line. 
In addition we study the more intense
lines Fe~{\sc i} $\lambda$6663.4~\AA\ and Ca~{\sc i} $\lambda$6717.7~\AA. 
Other intense lines included in different spectral orders were also measured.
All are neutral lines and, as the Li~{\sc i} line, the EW should be enhanced 
in a different way depending of their 
excitational potential when the the temperature decrease.
These lines are listed in Table~\ref{tab:linesp} together with their
excitational potential ($\chi$) and the mean EW ($\overline{\rm EW}$), 
the standard deviation ($\sigma$), and the
peak to peak variation (EW$_{\rm max}$ - EW$_{\rm min}$).
The measured EW for each night is plotted in Fig~\ref{fig:li_ews},
the errors bars have been calculated taking into account the 
error in placing of the continuum and the $\sigma$ obtained from 
repeated measures in each line.
As can be seen the variations in these lines are very small and,
contrary to the Li~{\sc i} line no correlation with the temporal evolution 
of the flare is evident.
The peak to peak variation in the Li~{\sc i} line is a factor 3 larger 
than in the other lines and the $\sigma$ of similar strength lines 
is 3 times smaller.  
Furthermore, no clear systematic behavior is observed 
with different excitation potentials.

%\placefigure{fig:li_ews}
%
%%+++++++++++++++++++
%%+++++++++++++++++++

\section{Discussion and conclusions}
%.................................................

%\subsubsection{Line blends?}
%.................................................

One source of variability of the
Li~{\sc i} $\lambda$6708~\AA\ line could due to  
blending with TiO bands at
6707.29, 6707.92, and 6708.16 \AA\ and CN bands at 6707.64 \AA\
as these molecular bands also become stronger at the lower temperatures
of the starspots.
However, the calculations of the Li~{\sc i} abundance in sunspots
that take into account these bands
(Engvold et al. 1970; 
Ritzenhoff et al. 1997)
concluded that the molecular blend is of lower importance.
Thus it seems reasonable assume that this effect in the stellar spectra
is also negligible.
At this spectral resolution the Li~{\sc i} line is blended with 
the nearby Fe~{\sc i} $\lambda$6707.41~\AA\ line, which 
%the EW of this line is normally 
%subtracted from the measured Li~{\sc i} + Fe~{\sc i} in order to obtain 
%the real value of the Li~{\sc i}  EW.
is clearly seen in the spectra of the inactive and Li free star
HR 5340 (K1~III). 
The mean EW measured in these spectra is 20~m\AA,
which is much smaller than the observed Li~{\sc i} EW.
Furthermore, as the other photospheric lines 
we measured did not show significant variations, we conclude that this 
line does not produce the variation in the measured Li~{\sc i}  EW.   
Due to the SB2 nature of this binary in some orbital phases the lines of
the primary component could be blended with the lines from the secondary
(see Fig.~\ref{fig:rej0743_li}). 
This is clearly seen in the spectrum from the 5th night (which is very
close the conjunction). The EW of the strongest lines measured during this
night are noticeably larger than in the rest of the nights due to the
contribution ($\approx$~15\%) to the EW from the secondary.
We have corrected the EW for this effect by subtracting the EW
of the lines in the secondary measured at other orbital phases where the
lines are not blended.
These corrected values are what we have plotted in Fig~\ref{fig:li_ews}.
However, for weak lines with EW similar to the Li~{\sc i} line
the contribution of the secondary seems to be negligible and this effect is
not observed.
In conclusion the observed Li~{\sc i} line variations do not seem to be
due to any kind of line blends.

%\subsubsection{Starspots and faculae?}
%.................................................

The Li~{\sc i} line variations could be caused by possible
cold spots and faculae on the stellar surface (see review by Fekel 1996).
Since this line is very temperature sensitive, 
the EW should be enhanced in dark
spots but reduced in the bright facular regions as shown by solar observations
(Grevesse 1968; Traub \& Roesler 1971; Giampapa 1984).
While Giampapa (1984) suggests this can substantially alter the EW in stellar 
spectra, other authors find this is not the case.
No detection of Li~{\sc i} EW variations in six active dwarfs 
have been reported by Boesgaard (1991).
The calculations of Soderblom et al. (1993) indicate that the effect
is only significant when the fraction of the surface covered by spots is 
very high (see also Stuik et al. 1997).
Pallavicini et al. (1993) show by means of spectral synthesis simulations 
that the effects may be less pronounced than suggested by Giampapa (1984)
and found no evidence that changes in the EW is
correlated with the photometric variability due to starspots in four active
stars.
The simulations done by Barrado (1996) also indicated smaller 
changes in the EW and even, in certain cases, the presence of faculae can
cancel these changes, leaving the EW unaltered.
By application of the Doppler imaging technique, Hussain et al. (1997) 
show that the Li~{\sc i} behaves in much the same way that conventional 
Doppler imaging Ca~{\sc i} and Fe~{\sc i} lines do.  
%These authors also found no evidence for the Li~{\sc i} abundance being 
%enhanced or depleted in starspots, 
%although small changes in the EW are possible, which can be diluted if
%spots were widely distributed in longitude.
% 
Until now     
significant variations in the Li~{\sc i} EW have been found only in 
some stars with very high Li~{\sc i} abundances such as 
pre-main sequence stars (Patterer et al. 1993; 
Fern\'{a}ndez \& Miranda 1998; Neuh\"auser et al. 1998)
and other young and very active stars 
(Robinson et al. 1986; 
Jeffries et al. 1994; Soderblom et al. 1996).
Large Li~{\sc i}  EW variation has been observed at
larger V band amplitude in V410 Tau.  
Here a peak to peak variability the Li~{\sc i} EW of 0.12~\AA\ 
has been found by Fern\'{a}ndez \& Miranda (1998) when the amplitude
in V was 0.6~mag, while little or no variation have been reported by
previous work on this star at lower V amplitudes
(Basri et al. 1991; Patterer et al. 1993; Mart\'{\i}n 1993;
Welty \& Ramsey 1995).
This result is confirmed on the young star Par~1724 by
Neuh\"auser et al. (1998).
However, in CAB little or no variations
have been previously reported (Pallavicini et al. 1993). 
Recent observations (Berdyugina et al. 1998)
of the extremely CAB
II Peg, which exhibits high V band variations and spot filling factors,
show very small Li~{\sc i} EW variations (10~m\AA), poorly correlated
with quasi-simultaneous photometric observations.
%.......
The Li~{\sc i} EW variations that we observe are clearly larger than 
those reported in other CAB with similar activity levels 
and Li~{\sc i} abundance.
Indeed, in other stars that exhibit large Li~{\sc i} EW variations  
other photospheric lines exhibit similar EW variations
(Fern\'{a}ndez \& Miranda 1998),
contrary to the behavior we report here.
Taking into account all these facts, the starspots that we 
infer on the surface of 2RE~J0743+224 from
the analysis of the TiO 7055~\AA\ band (Montes \& Ramsey 1998b), 
do not seem to be the primary cause
of the  observed Li~{\sc i} line variation.

%\placetable{tab:li}

%\placetable{tab:linesp}

%----------------------------------------------------------------
\begin{table}
\caption{Measured Li~{\sc i} $\lambda$6708~\AA\ line parameters
 \label{tab:li}}
\begin{center}
\scriptsize
\begin{tabular}{cccccc}
\noalign{\smallskip}
\hline
\noalign{\smallskip}
HJD          & EW & I &
$\Delta$ Ca - Li & $\lambda_0$ & $^6$Li/$^7$Li \\
(2450000+)   & (\AA) &   & (\AA) & (\AA) &   \\
\noalign{\smallskip}
\hline
\noalign{\smallskip}
 826.8794     & 0.102 & 0.146 & 9.892 & 6707.789 & 0. \\
 827.8537     & 0.122 & 0.149 & 9.899 & 6707.782 & 0. \\
 828.8925     & 0.133 & 0.161 & 9.898 & 6707.783 & 0. \\
 829.8929     & 0.126 & 0.165 & 9.886 & 6707.795 & 0. \\
 830.9239     & 0.136 & 0.164 & 9.889 & 6707.792 & 0. \\
 831.9308     & 0.138 & 0.177 & 9.880 & 6707.801 & 0. \\
 832.9098     & 0.153 & 0.178 & 9.861 & 6707.820 & 0.06 \\
 835.8891     & 0.130 & 0.144 & 9.853 & 6707.828 & 0.11 \\
\noalign{\smallskip}
\hline
\end{tabular}
\end{center}
\end{table}
%----------------------------------------------------------------

%----------------------------------------------------------------
\begin{table}
\caption{Photospheric line parameters
 \label{tab:linesp}}
\begin{center}
\scriptsize
\begin{tabular}{lcccccccccccccc}
\noalign{\smallskip}
\hline
\noalign{\smallskip}
%Cabecera de la tabla
Line Id.(M) & $\lambda$ & $\chi$ &
$\overline{\rm EW}$ & $\sigma$ & EW$_{\rm max - min}$  \\
             & (\AA)     & (eV) &
(\AA) &       & (\AA)                   \\
\noalign{\smallskip}
\hline
\noalign{\smallskip}
Li~{\sc i} (1)   & 6707.8   & 0.000 & 0.130 & 0.014 & 0.051 \\
\noalign{\smallskip}
%                 &          &       &        &       \\
%
Ca~{\sc i} (18)  & 6462.566 & 2.523 & 0.344 & 0.008 & 0.024 \\
Ca~{\sc i} (18)  & 6471.660 & 2.526 & 0.192 & 0.006 & 0.017 \\
Ca~{\sc i} (1)   & 6572.781 & 0.000 & 0.173 & 0.007 & 0.023 \\
Ca~{\sc i} (32)  & 6717.685 & 2.709 & 0.225 & 0.009 & 0.024 \\
\noalign{\smallskip}
%                 &          &       &        &       \\
%
Fe~{\sc i} (268) & 6546.245 & 2.758 & 0.174 & 0.006 & 0.021 \\
Fe~{\sc i} (111) & 6663.446 & 2.424 & 0.180 & 0.005 & 0.016 \\
Fe~{\sc i} (268) & 6703.573 & 2.758 & 0.101 & 0.005 & 0.017 \\
Fe~{\sc i} (34)  & 6710.310 & 1.485 & 0.094 & 0.003 & 0.009 \\
\noalign{\smallskip}
%                 &          &       &        &       \\
%
Ni~{\sc i} (43)  & 6643.641 & 1.676 & 0.172 & 0.005 & 0.019 \\
\noalign{\smallskip}
\hline
\end{tabular}
\end{center}
\end{table}
%----------------------------------------------------------------

%\subsubsection{Spallation reactions?}
%.................................................

%
The possibility of detecting Li~{\sc i} abundance inhomogeneities 
resulting from spallation reactions in the solar photosphere 
have been discussed by Hultqvist (1974, 1977). 
Evidence for such Li formation have been found through the 
deexcitation line Li~(478~keV) that results 
from $\alpha$-$\alpha$ reactions. This line
has been detected by $\gamma$-ray spectral observations  of solar flares 
with OSO-7 (Chupp et al. 1973), SMM (Murphy et al. 1990) and   
Yohkoh (Yoshimori et al. 1994; Kotov et al. 1996).
Recent calculation of Li production in solar flares by Livshits (1997)
agree with $\gamma$ line observations and suggest that enhancement of Li,
especially in the intensity of the Li~{\sc i} $\lambda$6708 \AA\ line, 
should be observed in the
Sun and other active stars.
Evidence for a Li enhancement at one umbral position, during a solar flare
is reported by Livingston et al. (1997).
In other stars (including the UV Ceti flares stars) 
no evidences of production of Li by nuclear reactions
have been previously reported.
The possibility of Li production have been discussed only
in terms of the energy required (Ryter et al. 1970; Karpen \& Worden 1979), 
 as a possibility to explain the high Li abundances observed in CAB
(Pallavicini et al. 1992) and stars with high flare activity
(Mathioudakis et al. 1995), to explain the widespread
presence of Li in very cool dwarfs (Favata et al. 1996),
or the $^6$Li detection in Population II stars (Deliyanis \& Malaney 1995).

%\subsubsection{$^6$Li/$^7$Li ratio  enhancement}
%.................................................

Another signature of Li production from spallation reactions is
that the $^6$Li/$^7$Li isotopic ratio should increase.
The predicted $^6$Li/$^7$Li ratio for the Li produced by spallation is
$\approx$~0.4 (Audouze 1970) or $\approx$~0.5 
(Walker et al. 1985).
The ratio measured in the Sun is
between 0.01 and 0.04 (Traub \& Roesler 1971; 
M\"{u}ller et al. 1975).
In Population I stars it is $\leq$~0.04
(Andersen et al. 1984; 
Maurice et al. 1984; Rebolo et al. 1986; 
Pallavicini et al. 1987), 
and in Population II stars it is $\approx$~0.05 in the
 two halo stars in which $^6$Li has been positively detected, but 
still lower upper limits are found for a larger number of other stars.
(Smith et al. 1993; Hobbs \& Thorburn 1994, 1997).
In order to estimate the $^6$Li/$^7$Li in our high resolution spectra
we adopt a method used by Herbig (1964) based on the shift of the center
of gravity (cog) of the Li~{\sc i} blend toward longer wavelengths as the
 $^6$Li fraction increases. 
If each  Li~{\sc i} component is weighted by its {\it gf}-value, 
pure $^7$Li would produce a cog wavelength of 6707.8117~\AA\, while 
pure $^6$Li would be 6707.9713~\AA. For weak lines, intermediate
mixtures would yield a wavelength, $\lambda_0$, between the two isotopes 
that would be weighted by the $^6$Li/$^7$Li ratio as
$^6$Li/$^7$Li = ($\lambda_0$ - 6707.8117) / (6707.9713 - $\lambda_0$).  
To determine the value of $\lambda_0$ we have measured the difference
between the Li~{\sc i} and Ca~{\sc i} features 
($\Delta$ = Ca~{\sc i} - Li~{\sc i}) where we adopt a wavelength of
6717.681~\AA\ for the Ca~{\sc i} line.
We give these values and the corresponding $^6$Li/$^7$Li ratio obtained 
in Table~\ref{tab:li}.
The largest difference in 
the inferred Li~{\sc i}) wavelengths amounts to only 0.046 \AA, or 30\% of one 
resolution element, and the effects of the blended Fe~{\sc i}, CN, and TiO 
lines on these inferred, apparent wavelengths of the Li~{\sc i} line are 
uncertain.
In order to estimate the possible errors 
we have also measured this difference, $\Delta$, for other photospheric 
lines included in the same spectral order than the Li~{\sc i} feature.
%(see Table~\ref{tab:li} and Fig.~\ref{fig:li_ll}).
%In Fig.~\ref{fig:li_ll} we plot $\Delta$ for the Li~{\sc i} lines and 
%the other lines.
The other line $\Delta$'s do not show any trend during the 
observations and the $\sigma$ with respect to the mean value is 
$\approx$~0.008.
The Li~{\sc i} line $\Delta$'s shows a tendency to decrease toward
the end of the flare, attaining a maximum difference 0.05.
This significant change in $\Delta$ and thus in the $^6$Li/$^7$Li ratio
is consistent with increasing $^6$Li during the flare
as is predicted for the production of Li~{\sc i} by spallation
reactions.

The  Li~{\sc i} EW variations that we observe are clearly correlated
with the temporal evolution of the flare,
and large changes are observed in the core of the Li~{\sc i} line,
as predict the models of Li production in flares (Livshits 1997).
Thus taking into account that
the other possible causes of variability have been minimized above
we suggest that this Li~{\sc i} is produced by spallation reactions
in the flare.
The observed $^6$Li/$^7$Li ratio also support this hypothesis.
Neither the EW variation or the $^6$Li/$^7$Li variation are
entirely convincing by themselves but together they are strongly suggestive.
This is the first time  that such Li~{\sc i} enhancement associate
with a stellar flare is reported, and probably the long-duration of
this flare is a key factor for this detection.

%.................................
%\acknowledgments

\begin{acknowledgements}

This work was supported by the Universidad Complutense de Madrid
and the Spanish Direcci\'{o}n General de Investigaci\'{o}n
Cient\'{\i}fica y  T\'{e}cnica (DGICYT) under grant PB94-0263,  
and by National Science Foundation (NSF) grant AST~92-18008.
We thank the staff of McDonald observatory for their allocation of 
observing time and their assistance with our observations.

\end{acknowledgements}

%.................................

%***************************** Figures ****************************

\end{document}